\documentclass[12pt]{article}
\usepackage{amsmath}
\usepackage{amssymb}

\usepackage{graphicx}

\usepackage[numbers,compress]{natbib}
\begin{document}
\title{Isotropic dynamical black holes}

\author{Ion I. Cot\u aescu\\ {\small \it  West 
                 University of Timi\c soara,}\\
   {\small \it V. P\^ arvan Ave. 4, RO-1900 Timi\c soara, Romania}}
\maketitle

\begin{abstract}
A new type of dynamical black holes is defined in manifolds with flat space sections having the  asymptotic behaviour of spatially flat Friedmann-Lema\^ itre-Robertson-Walker (FLRW) space-times.  These black holes are no longer vacuum solutions of the Einstein equations but preserve the isotropy of the flat space sections of the asymptotic FLRW manifolds.

Pacs: 04.70.Bw
\end{abstract}


\section{Introduction}

The black holes are vacuum solutions of the Einstein equations, with or without cosmological constant  $\Lambda$, derived in static local charts (called here frames) \cite{BH}. On the other hand, in cosmology one considers  evolving space-times as, for example,  the  FLRW ones which are the most plausible models of our universe in various epochs of evolution. Thus we are forced  to describe an expanding universe populated by static black holes whose asymptotic behaviour is different from that of the FLRW space-times. For this reason  we would like to focus here on this discrepancy looking for new metrics describing dynamical black holes with FLRW asymptotic behaviour.

As the actual universe is observed as being  spatially flat with a reasonable accuracy, we restrict ourselves to spatially flat FLRW space-times which have flat space sections. In these manifolds we can introduce physical frames, with coordinates of Painlev\' e - Gullstrand  type  \cite{Pan,Gul}, which have the advantage of laying out the apparent horizons that can give rise to some specific effects \cite{Kin,Cnew}. Obviously, the metrics we are looking for cannot be vacuum solutions of the Einstein equations in evolving backgrounds where the black holes must evolve accordingly.   Instead of the vacuum condition of the static black holes we require the dynamical ones to do not affect the isotropy of the flat space sections of the asymptotic background. In other words, the geometry of the dynamical black hole must preserve the space components of  the Einstein tensor of the asymptotic FLRW space-time.  These are the isotropic dynamic black holes whose metrics have to be studied here after a short review of the physical frames in which these may be defined. 

We use the Planck units with $\hbar=c=G=1$.

\section{Physical frames}

The black holes solutions are written in static  frames, $\{t_s, {\bf x}\}$, whose coordinates $x^{\mu}$ ($\alpha,\mu,\nu,...=0,1,2,3$) are the static time $t_s$ and Cartesian space coordinates  ${\bf x}=(x^1,x^2,x^3)$ associated to the spherical ones $(r,\theta,\phi)$. The geometry of the spherically symmetric black holes is defined by line elements of the form    
\begin{eqnarray}
ds^2=f(r)\, dt_s^2-\frac{dr^2}{f(r)}-r^2 d\Omega^2 \,,\label{s1s}
\end{eqnarray} 
where $d\Omega^2=d\theta^2+\sin^2\theta\, d\phi^2$. The function $f$ determines the properties of the black hole and the asymptotic form of the space-time.  For example, in the case of the Schwarzschild-de Sitter black hole,  the function
\begin{equation}\label{fr}
f(r)=1-\frac{2M}{r}-\omega_{dS}^2 r^2\,,
\end{equation}  
depends on the black hole mass $M$ and the Hubble-de Sitter constant denoted here by $\omega_{dS}=\sqrt{\frac{\Lambda}{3}}$. When $\omega_{dS}=0$ the metric becomes asymptotic flat  describing a Schwarzschild  black hole in  Minkowski  space-time. Otherwise, for  $\omega_{dS}\not=0$, we say that the black hole stays in a static frame of the de Sitter space-time. However, physically speaking,  the de Sitter manifold is not a static space-time as its expanding portion is a model of expanding universe governed by the {\em cosmic}, or proper, time $t$ defined as 
\begin{equation}
t=t_s-\int dr \frac{\sqrt{1-f(r)}}{f(r)}\,.
\end{equation}    
Substituting $t_s\to t$  we obtain the {\em physical} frame $\{t,{\bf x}\}$, with  coordinates of   Painlev\' e - Gullstrand  type  \cite{Pan,Gul}, in which the static metric (\ref{s1s}) takes the form
\begin{eqnarray}
ds^2=f(r)dt^2+2\sqrt{1-f(r)}\,dtdr-dr^2 -r^2d\Omega^2\,,\label{ss}
\end{eqnarray}
laying out flat space sections. In this manner, we may introduce physical frames  in any isotropic manifold  with flat space sections.

This suggests us to chose spatially flat FLRW space-times as backgrounds of the black holes we intend to study here. These manifolds  have conformal Euclidean frames, $\{t_c,{\bf x}_c\}$, with conformal time $t_c$ and the co-moving Cartesian space coordinates  ${\bf x}_c$, whose line elements read
\begin{equation}\label{conf}
ds^2=a(t_c)^2\left(dt_c^2-d{\bf x}_c\cdot d{\bf x}_c\right)\,.
\end{equation}
Note that the conformal coordinates were proposed for the first time by Lema\^ itre \cite{Lemco} in de Sitter's universe. The physical coordinates may be introduced performing the substitution
\begin{equation}\label{subs2}
t_c=\int \frac{dt}{a(t)}\,,  \quad {\bf x}_c=\frac{{\bf x}}{a(t)}\,,
\end{equation}
leading to the new line elements
\begin{equation}\label{Pan}
ds^2=\left(1-\frac{\dot a(t)^2}{a(t)^2}\, {{\bf x}}^2\right)dt^2+2\frac{\dot a(t)}{a(t)}\, {\bf x}\cdot d{\bf x}\,dt -d{\bf x}\cdot d{\bf x}\,, 
\end{equation}
of the physical frames $\{t,{\bf x}\}$ of these manifolds. The function 
$a(t)=a[t_c(t)]$ is the usual FRLW scale factor while
\begin{equation}\label{Hub}
\frac{\dot a(t)}{a(t)}=\frac{1}{a(t)}\frac{d a(t)}{dt}=\frac{1}{a(t_c)^2}\frac{d a(t_c)}{dt_c}\,,
\end{equation}
is the Hubble function for which we do not use a special notation.  The inverse transformation $\{t,{\bf x}\}\,\to\,\{t_c,{\bf x}_c\}$ is obvious
\begin{equation}\label{subs1}
t=\int a(t_c)dt_c\,,  \quad {\bf x}=a(t_c){\bf x}_c\,,
\end{equation}
As the geometry of these FLRW space-times is determined completely by the scale factors $a(t)$  we denote them by $(M,a)$ reminding the reader that their Einstein tensors have the components \cite{BH}
\begin{eqnarray}
G^0_{\, \, 0}(a) &=&3\,\frac{\dot{a}^2}{a^2}\,,\label{1}\\
G^i_{\, j}(a) &=&\delta^i_j \left(2\frac{\ddot{a}}{a} + \frac{\dot{a}^2}{a^2}\right)\,, \quad i, j,...=1,2,3\,,\label{2}
\end{eqnarray}
generating the Friedmann equations
\begin{eqnarray}
3\left(\frac{\dot a}{a}\right)^2&=&\Lambda + 8\pi \,\varepsilon\,, \label{E1}\\
2\frac{\ddot a}{a}
+\left(\frac{\dot a}{a}\right)^2 &=&\Lambda - 8\pi \, p \,,\label{E2}
\end{eqnarray}
of the model of an isotropic ideal gas with energy density $\varepsilon$ and pressure  $p$.  

The physical frame with the metric (\ref{Pan}) is the proper frame of an observer staying at rest in origin (${\bf x}=0$) and evolving along the vector field $\partial_t$ which, in general, is not a Killing one. In this frame the observer has  an evolving apparent horizon of radius
\begin{equation}
r_a(t)=\frac{a(t)}{\dot a(t)}\,.
\end{equation}
Moreover, this can have  an event horizon whose radius is the distance   \cite{R1,R2},    
\begin{equation}\label{evhor}
r_e(t)=\int_{t}^{\infty}dt'\frac{a(t)}{a(t')}\,,
\end{equation}
from which a photon emitted at the time $t$ never arrives in origin. In general, these horizons are different  evolving with different velocities \cite{Cnew}. The exception is the de Sitter space-time whose scale factor $a_{dS}$ determines a static event horizon which coincides with the apparent one,
\begin{equation}\label{adS}
a_{dS}(t)=e^{\omega_{dS}t}  ~~~\to~~~ r_a=r_e=\frac{1}{\omega_{dS}}\,.
\end{equation}
Thus in physical frames we have the opportunity of analysing the physical effects due to the apparent horizons  \cite{Cnew} which are hidden in the frames with co-moving space coordinates. In contrast, the event horizons which separate different domains of causality can be identified in any  frame where their role can be pointed out.

\section{Dynamical black holes}

For finding new objects in physical frames   we start with the simplest choice  substituting 
\begin{equation}
f(r)~~\to~~f(t,r)=1-\frac{2M}{r}-\frac{\dot{a}(t)^2}{a(t)^2}\,r^2\,,
\end{equation}
in Eq. (\ref{ss}). We obtain thus a new metric for which we derive the Einstein tensor whose structure is anisotropic. Therefore,  we have to look for particular scale factors  assuring the space isotropy compatible with the model of the ideal gas. We find that the unique solution is the de Sitter scale factor (\ref{adS}) which means that there are no metrics of this form describing other isotropic objects apart from the Schwarzschild-de Sitter black holes.

Under such circumstances, we focus on a new class of objects with spherical symmetry whose geometries are defined in the physical frames by the family of metrics
\begin{equation}\label{fam}
ds^2=\left[1-h_{\kappa}(t,r)^2\right]dt^2+2 h_{\kappa}(t,r)dr dt -dr^2-r^2 d\Omega^2\,,
\end{equation}
depending on the free parameter $\kappa\in {\Bbb R}$ as 
\begin{equation}\label{lineBH}
h_{\kappa}(t,r)=\frac{\dot{a}(t)}{a(t)}\,r+\sqrt{\frac{2 M_{\kappa}(t)}{r}}\,,\quad  M_{\kappa}(t)=M_0 \frac{ a(t_0)^{\kappa}}{a(t)^{\kappa}}\,.
\end{equation}
$M_{\kappa}(t)$ plays the role of a dynamical mass satisfying the initial condition $M_{\kappa}(t_0)=M_0$ at the initial time $t_0$.  We observe that for $r\to \infty$ all these metrics take the form (\ref{Pan}) such that we can say that our objects are evolving in  space-times  $(M,a)$. In other respects it is obvious that in the static flat case, when $a(t)=1$ and $\dot{a}(t)=0$, we recover the proper physical frame of a Schwarzschild black hole. 

It remains to convince ourselves that, in the general case of any scale factors,  these objects behave as dynamical black holes. For this purpose we study the ruts of the equation $g_{00}=0 \to h_{\kappa}(t,r)=\pm1$ in a neighbourhood of an initial time $t_0$ where we may  resort to the approximation
\begin{equation}\label{ap}
\frac{a(t)}{a(t_0)}\simeq 1+\omega (t-t_0)\,, \quad \omega=\left.\frac{\dot{a}(t)}{a(t)}\right|_{t=t_0}\,,
\end{equation}
assuming that $t_0$ is chosen such that  $\omega$ be very small as  $\omega^{-1}$, which is  the radius of the asymptotic apparent horizon, be of the order of the universe width. Then  the ratio between the Schwarzschild radius and  the  horizon one is also very small,  $2M_0\omega\ll 1$.   In this manner  we find four real  approximative solutions of the equation $g_{00}=0$ that read
\begin{eqnarray}
r_{b}(t)&=&2M_0\left[ 1+\kappa\omega(t-t_0) -4M_0\omega+{\cal O}(\omega^2) \right]\,,\label{rS}\\
r_{c}(t)&=&\frac{1}{\omega}\left[1-\sqrt{2M_0\omega}-M_0\omega  -\frac{1}{2}\kappa \omega\sqrt{2M_0\omega}\,(t-t_0) \right.\nonumber\\
&&\left.~~~~~~~~ - \frac{5}{4}\omega M_0 \sqrt{2M_0\omega}+{\cal O}(\omega^2)\right]\,,\label{rpm}
\end{eqnarray}
while other two complex valued solutions does not have a physical meaning. The  solution (\ref{rS}) plays the role of a dynamical Schwarzschild, or black hole,  radius which in the flat limit $\omega\to 0$ becomes the genuine one, $2M_0$. The solution (\ref{rpm}) define the cosmological  horizons that are close to the apparent horizon of radius $\omega^{-1}$ of the metric (\ref{Pan}). Thus we can conclude that our new objects can be interpreted as dynamical black holes in space-times $(M,a)$ at least in the time domains where we can use the approximation (\ref{ap}).   

It remains to see if there exists a convenient dynamical black hole with isotropic Einstein tensor. After a few manipulation we find that in a manifold having a physical frame with metrics (\ref{fam}) its components read 
\begin{eqnarray}
G^0_{\, \, 0} &=&3\,\frac{\dot{a}^2}{a^2} +\delta_{\kappa}\,,\\
G^r_{\, r} &=& 2\frac{\ddot a}{a}
+\left(\frac{\dot a}{a}\right)^2+\left( 1-\frac{\kappa}{3}\right)\delta_{\kappa}\,, \\
G^{\theta}_{\, \theta} =G^{\phi}_{\, \phi}& =& 2\frac{\ddot a}{a}
+\left(\frac{\dot a}{a}\right)^2+\frac{1}{4}\left( 1-\frac{\kappa}{3}\right)\delta_{\kappa}\,,
\end{eqnarray}
where
\begin{equation}
\delta_{\kappa}(t,r)= 3\frac{\dot a(t)}{a(t)}\sqrt{\frac{2 M_{\kappa}(t)}{r^3}}\,.
\end{equation}
Hereby we understand that for $\kappa=3$ we obtain the desired solution with isotropic Einstein tensor that can be used in cosmological models.

Summarising the previous  results we can say that we found a new type of dynamical black holes whose proper physical frames have the line elements 
\begin{equation}\label{fam1}
ds^2=\left[1-h(t,r)^2\right]dt^2+2 h(t,r)dr dt -dr^2-r^2 d\Omega^2\,,
\end{equation}
where now
\begin{equation}\label{lineBH1}
h(t,r)=h_{\kappa=3}(t,r)=\frac{\dot{a}(t)}{a(t)}\,r+\sqrt{\frac{2 M_0\, a(t_0)^3}{r\, a(t)^3}}\,.
\end{equation}
The corresponding Einstein tensors have isotropic components,
\begin{eqnarray}
G^0_{\, \, 0} &=&G^0_{\, \, 0}(a) +\delta\,,\\
G^i_{\, j} &=&G^i_{\, j}(a) \,,
\end{eqnarray}
where $G(a)$ is the Einstein tensor of the asymptotic manifold $(M,a)$ as given by Eqs. (\ref{1}) and (\ref{2}).  The presence of the dynamical black hole is encapsulated only in the additional  term
\begin{equation}
\delta(t,r)=\delta_{\kappa=3}(t,r)= 3\frac{\dot a(t)}{a(t)}\sqrt{\frac{2 M_{0}\,a(t_0)^3}{r^3\, a(t)^3}}\,,
\end{equation}
which vanishes in the flat limit when the black hole becomes a Schwarzschild one.

\section{Concluding remarks}

As the metrics of the genuine static black holes are vacuum solutions of the Einstein equations that can be derived exclusively in static manifolds, we propose a new type of dynamic black holes which  can be introduced in any spatially flat FLRW manifold $(M,a)$ without affecting dramatically the Friedmann equations. More specific, in a manifold generated by the gravitational sources $(\varepsilon,p)$ the black hole brings only the term which modify the density of energy, $\epsilon\to\epsilon+\delta/8\pi$, while the pressure remains unchanged preserving the isotropy of the space sections. The term $\delta/8\pi$ can be interpreted as the energy density of the coupling between the dynamical black hole and its background $(M,a)$.  

We have shown that the new dynamical black hole defined here behave close to the static ones in the time domains where the Hubble function is very small and the approximation (\ref{ap}) does hold. The difference is that the Schwarzschild and  apparent horizons are doubled  such that the role of these horizons must be studied in further investigations for identifying the possible event horizons similar to that of the static Schwarzschild-de Sitter black hole. Another problem is how these new objects behave in the space domains where the Hubble function takes larger values and the mentioned horizons are approaching or even overlapping each other. In these time domains we expect to find  new interesting properties that must be analysed carefully by using analytical and numerical methods.

Finally we note that we do not exclude the possibility of finding other types of isotropic dynamical black holes that could be used in various cosmological scenarios.

\end{document}